\documentclass[prl,aps,tightenlines,superscriptaddress,floatfix,footinbib]{revtex4-1} 
\usepackage{graphicx}
\usepackage{amsmath}
\usepackage{amsfonts,amsbsy}
\usepackage{amssymb}

\def\pasc{p^{\textrm{asc}}_T}
\def\ptrig{p^{\textrm{trig}}_T}
\def\as{\alpha_S}
\def\sp{S_\perp}
\def\qs{Q_S}
\def\Qs{\qs}
\def\nc{N_c}
\def\cf{C_F}
\def\qp{ {\bf q}_T } 
\def\pp{ {\bf p}_T } 
\def\kp{ {\bf k}_T } 
\def\qphat{ \hat{\bf q}_\perp } \def\pphat{ \hat{\bf p}_\perp }
\newcommand{\kpn}[1]{ {\bf k}_{#1\perp} } 
\newcommand{\qpn}[1]{ {\bf q}_{#1\perp} }

\begin{document}

\title{Evidence for BFKL and saturation dynamics from di-hadron spectra at the LHC}

\author{Kevin Dusling}
\affiliation{Physics Department, North Carolina State University, Raleigh, NC 27695, USA
}
\author{Raju Venugopalan}
\affiliation{Physics Department, Brookhaven National Laboratory,
  Upton, NY 11973, USA
}

\begin{abstract}
We demonstrate that rapidity separated di-hadron spectra in high multiplicity proton-proton collisions at the LHC can be quantitatively described by a combination of BFKL and saturation dynamics. Based on these results, we predict the systematics of di-hadron spectra in proton-nucleus  collisions at the LHC.

\end{abstract}

\maketitle

Proton-proton collisions in the high energy Regge-Gribov asymptotics of QCD are
described by the exchange of ladder like emissions of gluons. These gluons
carry very small fractions $x$ of the longitudinal momenta of the colliding
protons but have fixed but sufficiently large transverse momenta such that the
coupling $\as$ governing their emission is weak. In these high energy
asymptotics, large logarithms of $x$ accompany every such emission;
multi-particle production is described by the Balitsky-Fadin-Kuraev-Lipatov
(BFKL) equation~\cite{Balitsky:1978ic,Kuraev:1977fs} which performs a resummation of $\as\ln(x)$ terms that appear at each rung of the QCD ladder.  The BFKL bremsstrahlung  evolution equation in $x$ can be  contrasted to the   Dokshitzer-Gribov-Lipatov-Altarelli-Parisi (DGLAP) equations~\cite{Dokshitzer:1977sg,Gribov:1972ri,Altarelli:1977zs} corresponding to QCD evolution with squared momentum resolution $Q^2$. An important aspect of a quantitative description of high energy scattering in QCD is the regime of applicability of BFKL and DGLAP ladder resummations. 

Another important feature of QCD in Regge-Gribov asymptotics is gluon 
saturation~\cite{Gribov:1984tu,Mueller:1985wy}, which predicts that the
non-linear dynamics of gluons slows the growth of gluon distributions at
small $x$ due to maximal phase space occupancies of gluons for transverse momenta below the saturation scale $\Qs(x)$ in hadron
wavefunctions. This semi-hard scale is significantly larger at small $x$ than the intrinsic QCD scale, thereby making feasible a weak coupling treatment of the non-perturbative dynamics of QCD saturation~\cite{MV}. In small $x$ asymptotics,  both the growth of gluon distributions via BFKL evolution and the onset and properties of the saturation regime are described by the Color Glass Condensate effective field theory (CGC-EFT)~\cite{Gelis:2010nm}.  

In this letter, we will demonstrate that the combination of BFKL and saturation
dynamics in the leading power counting of the CGC-EFT provides a good description
of high multiplicity di-hadron spectra in proton-proton (p+p) collisions at the LHC. These spectra
were measured by the CMS collaboration~\cite{Khachatryan:2010gv} in events with
a high charged hadron multiplicity ($N>110$) trigger in the  rapidity range of
$2\leq |\Delta \eta| \leq 4$. On the nearside (with azimuthal separations
$\Delta \phi\approx 0$), the di-hadron spectra display a novel and mostly
unanticipated collimation called the ``ridge". In a previous letter~\cite{Dusling:2012iga}, we showed that the di-hadron yield in the p+p
ridge  can be explained by ``Glasma graphs"~\cite{Dumitru:2008wn,Dusling:2009ni,Dumitru:2010iy}: these graphs are
enhanced in high multiplicity events relative to minimum bias by
$\alpha_S^{-8}$, a factor of $10^4$--$10^5$ for typical values of $\alpha_S$. We will elaborate significantly on this study here. 

\begin{figure}[t]
\centering
\includegraphics[scale=1]{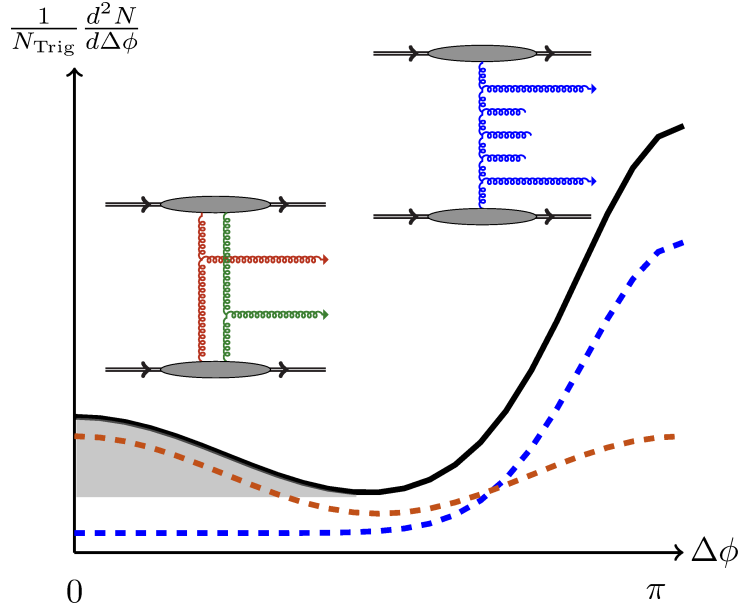}
\caption{Anatomy of di-hadron correlations.  The glasma graph on the left illustrates its 
its schematic contribution to the double inclusive cross-section (dashed orange curve). On the right is the 
back-to-back graph and the shape of its yield (dashed blue curve). The grey blobs denote emissions all the way from beam rapidities to 
those of the triggered gluons. The solid black curve represents the sum of contributions from glasma and back-to-back graphs. The shaded region represents the Associated Yield (AY) calculated using the zero-yield-at-minimum (ZYAM) procedure.}
\label{fig:graph}
\end{figure}

On the awayside ($\Delta \phi\approx \pi$), there is a small collimated Glasma graph contribution, but it is dominated by the better known  ``back-to-back" QCD
graphs. We will show that gluon emissions between $\pasc$ and $\ptrig$ partons are essential and can be described by a universal BFKL Green function.
Both Glasma and BFKL  graphs are illustrated in fig.~\ref{fig:graph}. Detailed exploration of di-hadron spectra has the potential to pin down, uniquely thus far in hadron-hadron collisions, the underlying perturbative ``QCD string" dynamics, gluon saturation, and the interplay between the two.  In this spirit, we make predictions for di-hadron spectra in both minimum bias and high multiplicity proton-Lead (p+Pb) $\sqrt{s}=5.02$ TeV collisions at the LHC. 

We first examine the Glasma Graphs for correlated two gluon production. In
Ref.~\cite{Dusling:2012iga}, we wrote down the leading contributions in
$\pp/\qs$, which can be expressed in terms of the two particle momentum space rapidities $y_{p,q}$ and transverse momenta $\pp,\qp$ as
 \begin{align}
\frac{d^2N_{\rm \sl Glasma}^{\rm \sl corr. (1) }}{d^2\pp d^2\qp dy_p dy_q}
= \frac{C_2}{\pp^2\qp^2} \int_{\kp} (D_1 + D_2) \, ,
\label{eq:Glasma-corr1}
\end{align}
with $C_2 = \frac{\alpha_S(\pp)\,\alpha_S(\qp)}{4\pi^{10}}\frac{N_C^2 \, S_\perp}{(N_C^2-1)^3 \,\zeta}$ and
\begin{align}
D_1 &= \Phi_{A_1}^2(y_p,\kp)\Phi_{A_2}(y_p,\pp-\kp)
D_{A_2}\nonumber \\
D_2 &= \Phi_{A_2}^2(y_q, \kp)\Phi_{A_1}(y_p,\pp-\kp)
D_{A_1}\, ,
\end{align}
where  $D_{A_{2(1)}} =
\Phi_{A_{2(1)}}(y_q,\qp+\kp)+\Phi_{A_{2(1)}}(y_q,\qp-\kp)$.  We extend 
our previous calculation by evaluating additional diagrams that are formally subleading in
$\pp/\qs$, as listed in appendix A of Ref.~\cite{Dusling:2009ni}. The
sub-leading contributions discussed in that work\footnote{The different color topologies that correspond to the leading and sub-leading contributions are discussed in detail in Ref.~\cite{Dumitru:2008wn}; in particular, see Figs. 7 and 8, and 
accompanying text.} are necessary for a more quantitative description of 
data. The contribution of these additional diagrams can be expressed as  
\begin{align}
\frac{d^2N_{\rm \sl Glasma}^{\rm \sl corr. (2) }}{d^2\pp d^2\qp dy_p dy_q} = \frac{C_2}{\pp^2\qp^2} \sum_{j=\pm}\left(A_1(\pp,j\qp) + \frac{1}{2} A_2(\pp,j\qp)\right)\,,
\label{eq:Glasma-corr2}
\end{align}
with\footnote{The delta function $\delta(\pp\pm\qp)$ in $A_1$ is broadened by multiple
scattering as well as fragmentation effects the full treatment of which is beyond our scope. Here we smear the distribution to be $\delta(\phi_{pq})\to
\frac{1}{\sqrt{ 2\pi \sigma}}\exp\left(-\phi_{pq}^2/2\sigma^2\right)$, where $\Delta\phi_{p,q}=\phi_p-\phi_q$ and $\sigma=3\textrm{ GeV}/\pp$
is a $\pp$ dependent width on the order of the saturation scale.  We stress that the associated yield--the integral over the near-side
signal--is insensitive to details of this smearing.}
$A_1 = \delta^2(\pp+\qp) \left[\mathcal{I}_1^2 + \mathcal{I}_2^2 + 2\mathcal{I}_3^2 \right]$, such that 
\begin{align}
\mathcal{I}_1&=\int_{\kpn{1}} \Phi_{A_1}(y_p,\kpn{1})
\Phi_{A_2}(y_q,\pp-\kpn{1}) \frac{\left(\kpn{1}\cdot
\pp-\kpn{1}^2\right)^2}{\kpn{1}^2\left(\pp-\kpn{1}\right)^2}\;,\nonumber\\
\mathcal{I}_2&=\int_{\kpn{1}}  \Phi_{A_1}(y_p,\kpn{1})
\Phi_{A_2}(y_q,\pp-\kpn{1})\frac{\left|\kpn{1}\times\pp\right|^2}{\kpn{1}^2\left(\pp-\kpn{1}\right)^2}\;,\nonumber\\
\mathcal{I}_3&=\int_{\kpn{1}}  \Phi_{A_1}(y_p,\kpn{1})
\Phi_{A_2}(y_q,\pp-\kpn{1})\frac{\left(\kpn{1}\cdot
\pp-\kpn{1}^2\right)\left|\kpn{1}\times\pp\right|}{\kpn{1}^2\left(\pp-\kpn{1}\right)^2}\;.\nonumber
\end{align}
%\begin{align}
%\mathcal{I}_1&=&\int_{\kpn{1}} {\cal F}^{(1)}(y_p,y_q,\kpn{1},\pp-\kpn{1})
%\frac{\left(\kpn{1}\cdot \pp-\kpn{1}^2\right)^2}{\kpn{1}^2\left(\pp-
%\kpn{1}\right)^2}\nonumber\\
%\mathcal{I}_2&=&\int_{\kpn{1}}  {\cal F}^{(1)}(y_p,y_q,\kpn{1},\pp-
%\kpn{1})\frac{\left|\kpn{1}\times\pp\right|^2}{\kpn{1}^2\left(\pp-
%\kpn{1}\right)^2}\nonumber\\
%\mathcal{I}_3&=&\int_{\kpn{1}}  {\cal F}^{(1)}(y_p,y_q,\kpn{1},\pp-
%\kpn{1})\frac{\left(\kpn{1}\cdot \pp-
%\kpn{1}^2\right)\left|\kpn{1}\times\pp\right|}{\kpn{1}^2\left(\pp-
%\kpn{1}\right)^2}\nonumber\\
%\end{align}
%and ${\cal F}^{(1)}(y_p,y_q,\kpn{1},\pp-\kpn{1})= \Phi_{A_1}(y_p,\kpn{1})
%\Phi_{A_2}(y_q,\pp-\kpn{1})$. 
The other contribution, $A_2$ in Eq.~(\ref{eq:Glasma-corr2}) can be expressed as 
\begin{eqnarray}
A_2 =&& \int_{\kpn{1}}
\Phi_{A_1}(y_p,\kpn{1})\Phi_{A_1}(y_p,\kpn{2})
\Phi_{A_2}(y_q,\pp-\kpn{1})
\Phi_{A_2}(y_q,\qp+\kpn{1})
\nonumber\\
&\times&\frac{\left(\kpn{1}\cdot \pp-\kpn{1}^2\right)\left(\kpn{2}\cdot \pp-
\kpn{2}^2\right)+
\left(\kpn{1}\times\pp\right)\left(\kpn{2}\times\pp\right)}{\kpn{1}^2\left(\pp-
\kpn{1}\right)^2}\nonumber\\
&\times&\frac{\left(\kpn{1}\cdot \qp-\kpn{1}^2\right)\left(\kpn{2}\cdot \qp-
\kpn{2}^2\right)+
\left(\kpn{1}\times\qp\right)\left(\kpn{2}\times\qp\right)}{\kpn{2}^2\left(\qp+
\kpn{1}\right)^2}
\label{eq:double-inclusive-5}
\end{eqnarray}
where $\kpn{2}\equiv \pp-\qp-\kpn{1}$. 
%\begin{align}
%A_2 = \int_{\kpn{1}}{\cal F}^{(2)}(y_p,y_q,\kpn{1},\kpn{2},\pp-\kpn{1},\qp+\kpn{1}) {{\cal A}(\pp,\kpn{1},\kpn{2}) {\cal A}(\qp,\kpn{1},\kpn{2})
%\over {\kpn{1}^2\left(\pp-\kpn{1}\right)^2} {\kpn{2}^2\left(\qp+\kpn{1}\right)^2}} \,,
%\end{align}
%with ${\cal A}(\pp,\kpn{1},\kpn{2}) = \left(\kpn{1}\cdot \pp-\kpn{1}^2\right)\left(\kpn{2}\cdot \pp-\kpn{2}^2\right)+ \left(\kpn{1}\times\pp\right)\left(\kpn{2}\times\pp\right)$, $\kpn{2} = \pp-\qp-\kpn{1}$ and 
%\begin{align}
%{\cal F}^{(2)}(y_p,y_q,\kpn{1},\kpn{2},\pp-\kpn{1},\qp+\kpn{1})=\Phi_{A_1}(y_p,\kpn{1})\Phi_{A_1}(y_p,\kpn{2})
%\Phi_{A_2}(y_q,\pp-\kpn{1})
%\Phi_{A_2}(y_q,\qp+\kpn{1}) \, .\nonumber
%\end{align}

%
%%%%%%%%%%%%%%%%%%%%%%%%%%%%%%%%%%%%%%%%%%%%%%%%%%%%%%%%%%%%%%%%%%%%%%%%%
%%%%%%%%%%%%%%%%%%%%%%%%%%%%%%%%%%%%%%%%%%%%%%%%%%%%%%%%%%%%%%%%%%%%%%%%% 

%
%For our computation, we will also need the single inclusive gluon distribution
%\begin{align}
%\frac{\ud N_1}{\ud y_p\ud^2\pp }
%=\frac{C_1}{\pp^2}
%\int_{\kp}\!\!\!
%\Phi_{A_1}(y_p,\kp)\Phi_{A_2}(y_p,\pp-\kp)\,,
%\label{eq:single-incl}
%\end{align}
%with the coefficient $C_1=\frac{\alpha_s N_C S_\perp}{4\pi^6 (N_C^2-1)}$. 

The unintegrated gluon distribution (UGD) per unit transverse area $\Phi$ in these
expressions is a universal quantity determined by solving the
Balitsky-Kovchegov (BK) equation~\cite{Balitsky:1995ub,Kovchegov:1999yj} as a
function of the rapidity $y=\log\left(x_0/x\right)$. To avoid repetition, we
refer the interested reader to Ref.~\cite{Dusling:2012iga} for identical
details\footnote{Variations relative to Ref.~\cite{Dusling:2012iga} are as follows: a) we take $\as$ to run as a function of $\pp$ or $\qp$ instead of the saturation scale $\qs$. 
b) We use the NLO KPP  parametrization~\cite{Kniehl:2000fe} for gluon fragmentation 
to charged hadrons. c) The scale in the BK equation at the initial value $x=x_0$ is adjusted slightly
to be $Q_0^2=0.168\textrm{ GeV}^2$ for min bias p+p collisions and
$Q_0^2=0.672\textrm{ GeV}^2$ for central p+p collisions--we emphasize, as in
Ref.~\cite{Dusling:2012iga}, that this scale should not be confused with the much larger 
saturation scale $\qs$.} of i) the derivation of Eqs.~(\ref{eq:Glasma-corr1})
and (\ref{eq:Glasma-corr2}),  ii) the expression for the single inclusive distribution, iii) solutions of the BK equation,  iv) a discussion of the non-perturbative constant $\zeta$ in $C_2$ that represents the effect of soft multi-gluon interactions on the di-hadron spectrum. 

We now turn to the double inclusive distribution from the back-to-back QCD
graphs shown in Fig.~\ref{fig:graph}. At high energies, for a pair of hadrons
having a rapidity separation $\Delta y\gtrsim 1/\alpha_s$, a resummation of
rapidity ordered multi-gluon emissions is necessary, corresponding to
$t$-channel Pomeron exchange in the language of ``Reggeon Field
Theory"~\cite{Lipatov:1995pn,DelDuca:1995hf}.  These emissions between the two
tagged partons lead to an angular decorrelation of the dihadron signal observed
in the data.  The observation of an angular decorrelation as a signal of pomeron exchange is complementary to looking for the growth in the dijet cross-section, as first proposed by Mueller and Navelet \cite{Mueller:1986ey}. In this framework, the double inclusive multiplicity can be expressed as~\cite{Colferai:2010wu,Fadin:1996zv}
\begin{eqnarray}
\label{eq:BFKL}
\left.\frac{d^2N_{AB}}{d^2\pp d^2\qp dy_p dy_q}\right|_{\rm BFKL} & &= \frac{32\,\nc\,
\alpha_s(\pp)\,\alpha_s(\qp)}{ (2\pi)^8 \,\cf}\,\frac{\sp}{\pp^2\qp^2}\\
&\times&\int_{\kpn{0}} \int_{\kpn{3}}
\Phi_A(x_1,\kpn{0})\Phi_B(x_2,\kpn{3})\,\mathcal{G}(\kpn{0}-\pp,\kpn{3}+\qp,y_p-y_q)\nonumber
\end{eqnarray}
where $\mathcal{G}$ is the BFKL Green's function\footnote{
Appearing in the BFKL Green's function is the BFKL eigenvalue,
${\omega(\nu,n)=-2\overline{\alpha}_s\,
\textrm{Re}\left[\Psi\left(\frac{|n|+1}{2}+i\nu\right)-\Psi(1)\right]}$, where 
$\Psi(z)= d\ln\Gamma(z)/dz$ is the logarithmic derivative of the Gamma function,
$\overline{\alpha}_s\equiv
\nc\,\as\left(\sqrt{\qpn{a}\qpn{b}}\right)/\pi$ and 
$\overline{\phi}\equiv \arccos\left(\frac{\qpn{a}\cdot \qpn{b}}{\vert\qpn{a}\vert\textrm{ }\vert \qpn{b}\vert}\right)$. 
}
\begin{eqnarray}
\mathcal{G}(\qpn{a},\qpn{b},\Delta y)=\frac{1}{(2\pi)^2}\frac{1}{(\qpn{a}^2 \qpn{b}^2)^{1/2}}\sum_n e^{in\overline{\phi}}\int_{-\infty}^{+\infty} d\nu\textrm{ } e^{\omega(\nu,n)\Delta y}e^{i\nu\ln\left(\qpn{a}^2/\qpn{b}^2\right)}\textrm{   } \, .
\label{eq:BFKL-Green}
\end{eqnarray}

The $\Phi$'s in the derivation of Eq.~(\ref{eq:BFKL}) are known as impact factors and, to next-to-leading-logarithmic (NLLx) accuracy in $x$, satisfy an integral
 equation where the kernel is the real part of the leading-logs in $x$ (LLx) BFKL kernel~\cite{Fadin:1999qc,Bartels:2002yj,Colferai:2010wu}. Since the CMS di-hadron kinematics requires very significant evolution of the $\Phi$'s down from beam rapidities, well beyond the Mueller-Navelet regime, we will make the ansatz here  that the $\Phi$'s are equivalent to the unintegrated gluon distributions in Eqs.~(\ref{eq:Glasma-corr1}) and (\ref{eq:Glasma-corr2}). 

As a test of this ansatz, we can ``turn off" the gluon radiation between the
jets by taking the $\as\Delta y\to 0$ limit of the BFKL Green's function\footnote{In taking the limit of $\alpha_s\Delta y\to 0$ it is useful to use the 
following integral representation of the $\delta$-function; 
$\delta^2(\pp-\qp)= \frac{1}{2\pi^2 \left(\pp^2\qp^2\right)^{1/2}}\sum_{n=-\infty}^{+\infty}\int_{-\infty}^{+\infty}d\nu\; e^{i\nu\ln\left(\pp^2/\qp^2\right)} e^{in\;\arccos(\qphat\cdot\pphat)}$.}
and obtain for the di-hadron differential cross-section, the well known expression in Multi-Regge kinematics (MRK)~\cite{Fadin:1996zv,Leonidov:1999nc}, 
\begin{eqnarray}
\left.\frac{d^2N_{AB}}{d^2\pp d^2\qp dy_p dy_q}\right\vert_{\rm MRK} = 
\frac{16\,\nc\, \alpha_s(\pp)\,\alpha_s(\qp)}{ (2\pi)^8 \,\cf}\, \frac{\sp}{\pp^2\qp^2}\int_{\kpn{1}} \Phi_A(x_1,\kpn{1})\Phi_B(x_2,\kpn{2})\,.
\label{eq:MRK}
\end{eqnarray}

The universal BFKL Green's function in Eq.~(\ref{eq:BFKL-Green}) resums leading
logs in $x$, albeit with $\as$ chosen to run as the geometrical mean of the two momentum scales. NLLx expressions for this quantity are now known~\cite{Andersen:2003wy,Colferai:2010wu} 
and are significantly more analytically cumbersome than
Eq.~(\ref{eq:BFKL-Green}). The moments ${\cal C}_m = \langle \cos
m(\Delta\phi_{p,q})\rangle$ ($\langle \cdots\rangle$ denotes the $\frac{d^2N_{AB}}{d^2\pp d^2\qp dy_p dy_q}$ weighted average over $\phi_p$,$\phi_q$) have been computed numerically to estimate the relative LLx and NLLx contributions to the double inclusive distributions for relative rapidities $\Delta Y \geq 6$~\cite{Caporale:2012qd,Colferai:2010wu}. While the ratios ${\cal C}_{1,2}/{\cal C}_0$ are larger at NLLx by factors of 2-3 relative to LLx, this is largely because the $\Delta\phi_{p,q}$ independent background ${\cal C}_0$ is significantly lower at NLLx relative to LLx. In contrast, the collimated contributions ${\cal C}_1$ and ${\cal C}_2$ at NLLx are larger than the LLx values by $\sim 10$\% and $30$\% respectively. 

In this work, we are interested in contributions to di-hadron production that are collimated in $\Delta\Phi_{p,q}$, above the 
$\Phi_{p,q}$ pedestal. Thus while a full NLL treatment is desirable, the relatively small NLLx corrections to ${\cal C}_{1,2}$ suggest a LLx treatment may be adequate. Further, because we include the effect of of running coupling both in the $\Phi$'s and ${\cal G}$--a piece of the NLLx contribution-the contributions neglected may be even smaller and comparable to other uncertainties such as choices of energy and renormalization scales. 

\begin{figure}
\includegraphics[width=4in]{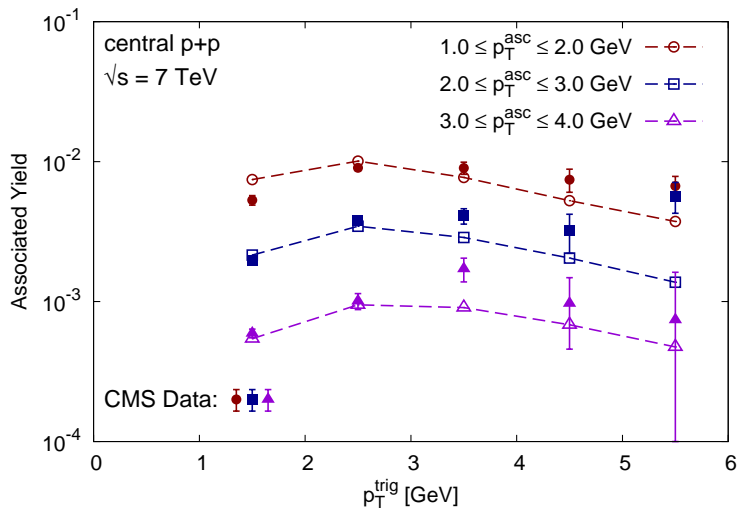}
\caption{The integrated associated nearside yield per trigger as a function of $\ptrig$
in three different $\pasc$ windows for high multiplicity p+p collisons.  Filled
symbols denote CMS data points extracted from the published $\ptrig$, $\pasc$ di-hadron matrix~\cite{Khachatryan:2010gv}. Open symbols are our results, with dashed lines between points to guide the eye.}
\label{fig:ay_centralpp}
\end{figure}

We now turn to a quantitative comparison of the Glasma graphs (summing contributions from Eqs.~(\ref{eq:Glasma-corr1}), 
(\ref{eq:Glasma-corr2})) and back-to-back graphs (Eq.~\ref{eq:BFKL}) to di-hadron spectra in high multiplicity collisions at the LHC. We first consider the 
associated yield per trigger\footnote{The transverse overlap area in the collision $S_\perp$ cancels in this ratio. Another key feature of the ZYAM procedure is that $\Delta \phi_{qp}$-independent 
contributions, which are known to receive contributions from multiple sources in
QCD~\cite{Dumitru:2010mv,Dumitru:2011zz,Kovner:2010xk,Kovner:2011pe}, do not
contribute.} shown in the shaded region of Fig.~(\ref{fig:graph}). The detailed
expressions for computing this yield implementing the ZYAM procedure are listed
in ~\cite{Dusling:2012iga}; as noted the novel feature here is the inclusion of
the contributions in Eq.~(\ref{eq:Glasma-corr2}). The results for the CMS high
multiplicity (denoted here as ``central p+p") events~\cite{Khachatryan:2010gv}
are shown in Fig.~(\ref{fig:ay_centralpp}).  The agreement with data is quite
good; the results in the $2 \leq \pasc \leq 3$ GeV and $3 \leq \pasc\leq 4$ windows have not been presented 
previously. While the trends are reproduced, data is slightly underestimated
with increasing $\ptrig$ (albeit within experimental uncertainties). We note
that in the absence of a Monte-Carlo generator, we are unable to require $N_{\rm
trig}\geq 2$, as done for the data, in each transverse momentum bin. Imposing
this cut would harden the spectrum in the correct direction.

%%%%%%%%%%%%%%%%%%%%%%%%%%%%%%%%%%%%%%%%%%%%%%%%%%%%%%%%%%%%%%%%%%%%%%%%%
%%%%%%%%%%%%%%%%%%%%%%%%%%%%%%%%%%%%%%%%%%%%%%%%%%%%%%%%%%%%%%%%%%%%%%%%%

\begin{figure}
\includegraphics[width=3in]{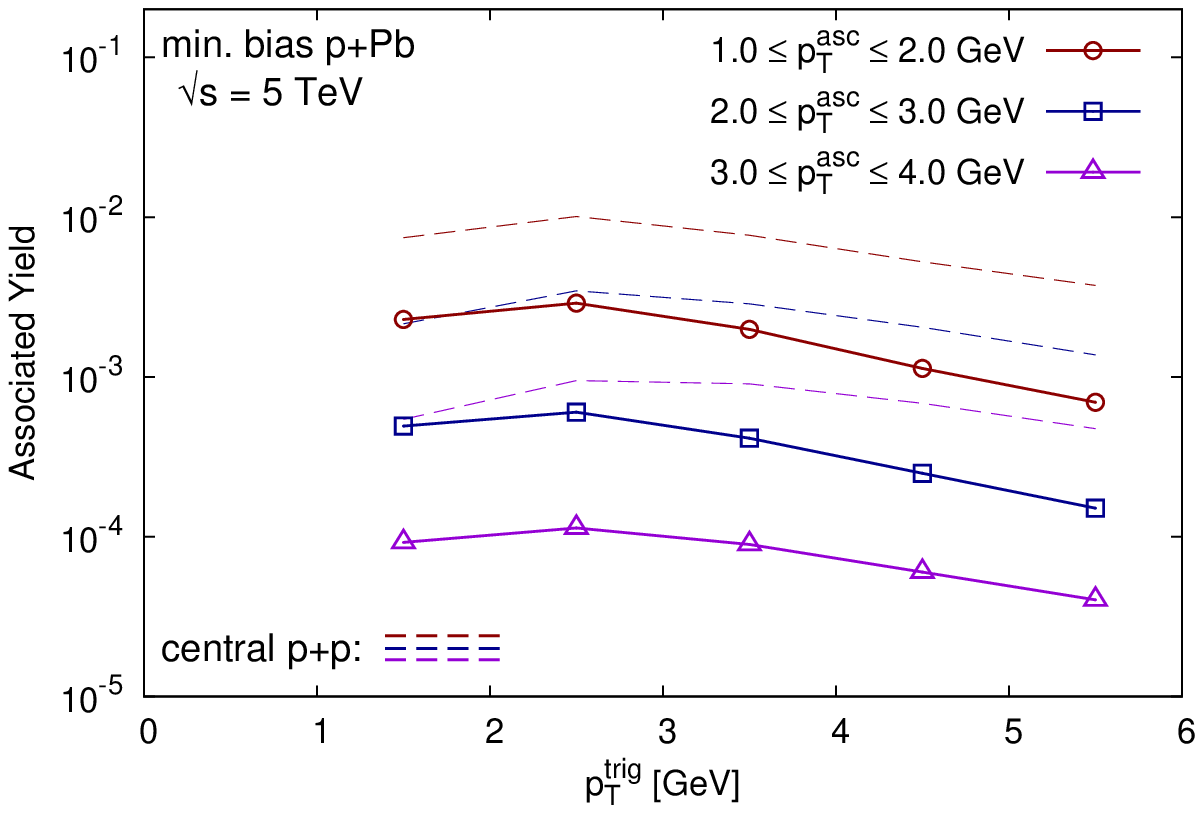}
\includegraphics[width=3in]{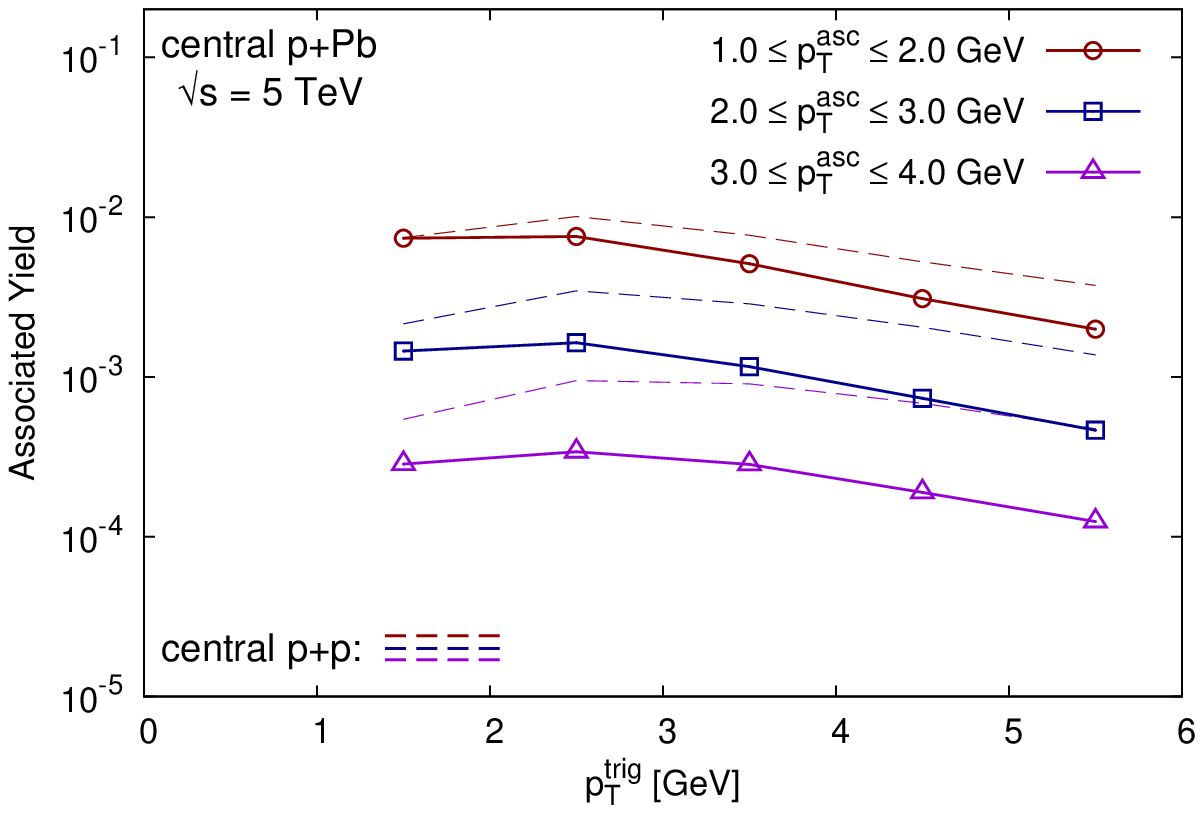}
\caption{The associated yield per trigger as a function of $\ptrig$ in min.bias p+Pb collisions at $\sqrt{s}=5$ TeV (left) and central p+Pb collisions (right). Predictions 
are for the  same $\pasc$ windows as for high multiplicity p+p collisons, and results from the latter are presented for comparison as dashed lines.}
\label{fig:ay_pPb}
\end{figure}

Inclusion of the contribution from Eq.~(\ref{eq:Glasma-corr2}) lowers the $K$-factor
from $K=2.3$ required previously in \cite{Dusling:2012iga} to $K=1$.  One important source of uncertainty is due to the non-perturbative constant\footnote{$\zeta$ is constrained by numerical computations~\cite{Schenke:2012hg} and fits to the n-particle p+p multiplicity distribution~\cite{Schenke:2012hg,Tribedy:2010ab,Tribedy:2011aa} to be $\zeta\sim 1/6$. However, these estimates are not definitive and there is considerable room for improvement.} $\zeta$.  In addition, an important caveat\footnote{\protect\label{anti-coll}This is practically the case (as we will discuss in detail); however, a very small anti-collimation on the nearside that is completely negligible relative to the awayside amplitude can skew the nearside {``ridge"} signal completely because it is so small in comparison. BFKL predictions are not of this 
accuracy--we will therefore conjecture that a more precise treatment of its
dynamics will give a $\Delta\phi_{pq}$ independent nearside contribution
{``within ridge accuracy"}.} is that we have taken the BFKL contribution on the
near-side to be independent of $\Delta \phi_{pq}$ for $\Delta\phi_{pq} \leq \pi/2$
in Figs.~(\ref{fig:ay_centralpp}) and (\ref{fig:ay_pPb}). 

One way of addressing these theoretical uncertainties is to make predictions for the associated per trigger yield for p+Pb collisions\footnote{Both Glasma and 
BFKL graphs were considered previously in ~\cite{Kharzeev:2004bw} to address forward-central di-hadron correlations in deuteron-gold collisions at RHIC. The relative power counting of the two contributions was however not well understood at that time. BFKL graphs were also discussed in the context of deuteron-gold 
collisions at RHIC in ~\cite{Tuchin:2009nf}.} and compare these to forthcoming LHC data on the same. In Fig.~(\ref{fig:ay_pPb}), we show predictions for the nearside ridge for both minimum bias (left) and central (right) p+Pb collisions. All parameters are fixed to be the same for p+p and p+Pb. 
The only difference\footnote{These values for $Q_0$ for the proton and Lead are (imperfectly) constrained by fits to available $p_T$ and rapidity distributions from RHIC and LHC for p+p and 
deuteron-gold collisions, and for the former from HERA e+p data as well. Once
data for these {``bulk"} quantities are available for p+Pb collisions, (very)
limited fine tuning of our results may be feasible.} is the value of the initial
scale $Q_0$ in the running coupling BK evolution of $\Phi$ for Pb nuclei.  The
initial saturation scale (in the fundamental representation) is taken to be
three times that of the min. bias proton value, $Q_0^2 =0.504$ GeV$^2$, and denotes what is meant in our context by ``minimum
bias" for lead ions. A value six times larger, $Q_0^2=1.008$ GeV$^2$, is used for ``central" ions.  Recall that the min. bias and ``central" p+p values were taken to be $Q_0^2=0.168$ GeV$^2$ and $Q_0^2=0.672$ GeV$^2$ respectively~\footnote{These initial values should not be confused with the saturation momentum, defined here as the peak of $\Phi(y,\kp)$, at the much smaller values of $x\sqrt{s}=p_T e^{\pm y}$ probed in the di-hadron studies here, which span (in the adjoint representation), $\qs^2\left(x\sim 10^{-4\div 5}\right)\sim 1-5$ GeV$^2$ respectively for initial saturation scales used in this work.}. The key message
from Fig.~(\ref{fig:ay_pPb}) is the prediction that there is a ridge in min.
bias p+Pb collisions at $\sqrt{s}=5$ TeV, albeit smaller than that in ``central"
p+p data. The ridge is larger in central p+Pb collisions\footnote{In this
context, one should note that the proton is always treated as min. bias in what
we have called {``central"} p+Pb; one should choose larger values for the initial
saturation scale for very high multiplicity p+Pb collisions where rare {``hot spot"}
configurations of the proton are selected.} but still smaller than central p+p
data. As for the latter, one expects a systematic under-prediction at the higher $p_{T,{\rm trig}}$ values due to the previously discussed $N_{\rm trig}\geq 2$ effect.

%%%%%%%%%%%%%%%%%%%%%%%%%%%%%%%%%%%%%%%%%%%%%%%%%%%%%%%%%%%%%%%%%%%%%%%%%
%%%%%%%%%%%%%%%%%%%%%%%%%%%%%%%%%%%%%%%%%%%%%%%%%%%%%%%%%%%%%%%%%%%%%%%%%

%%%%%%%%%%%%%%%%%%%%%%%%%%%%%%%%%%%%%%%%%%%%%%%%%%%%%%%%%%%%%%%%%%%%%%%%%
\begin{figure}[!ht]
\includegraphics[width=\textwidth]{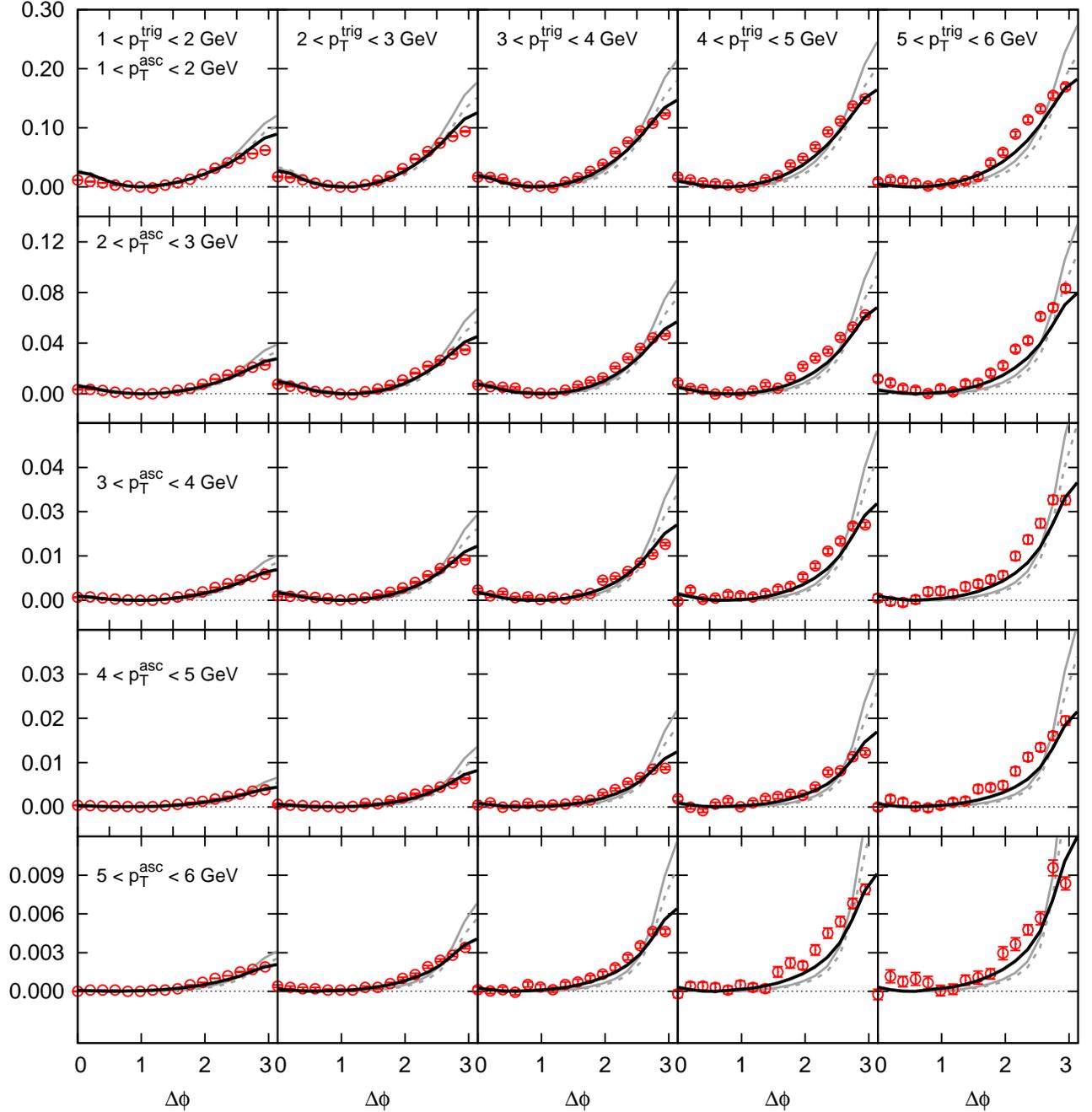}
\caption{CMS data after the removal of the underlying event in each bin via the ZYAM procedure.  The solid curve is the BFKL result added to the Glasma result
with K-factors of $K_{\rm{BFKL}}=1.1$ and $K_{\rm{Glasma}}=2.3$. The solid
(dashed) gray curves show the result without BFKL evolution in the MRK (QMRK)
framework.}
\label{fig:pp_matrix}
\end{figure}
%%%%%%%%%%%%%%%%%%%%%%%%%%%%%%%%%%%%%%%%%%%%%%%%%%%%%%%%%%%%%%%%%%%%%%%%%

%%%%%%%%%%%%%%%%%%%%%%%%%%%%%%%%%%%%%%%%%%%%%%%%%%%%%%%%%%%%%%%%%%%%%%%%%
\begin{figure}[!ht]
\includegraphics[width=\textwidth]{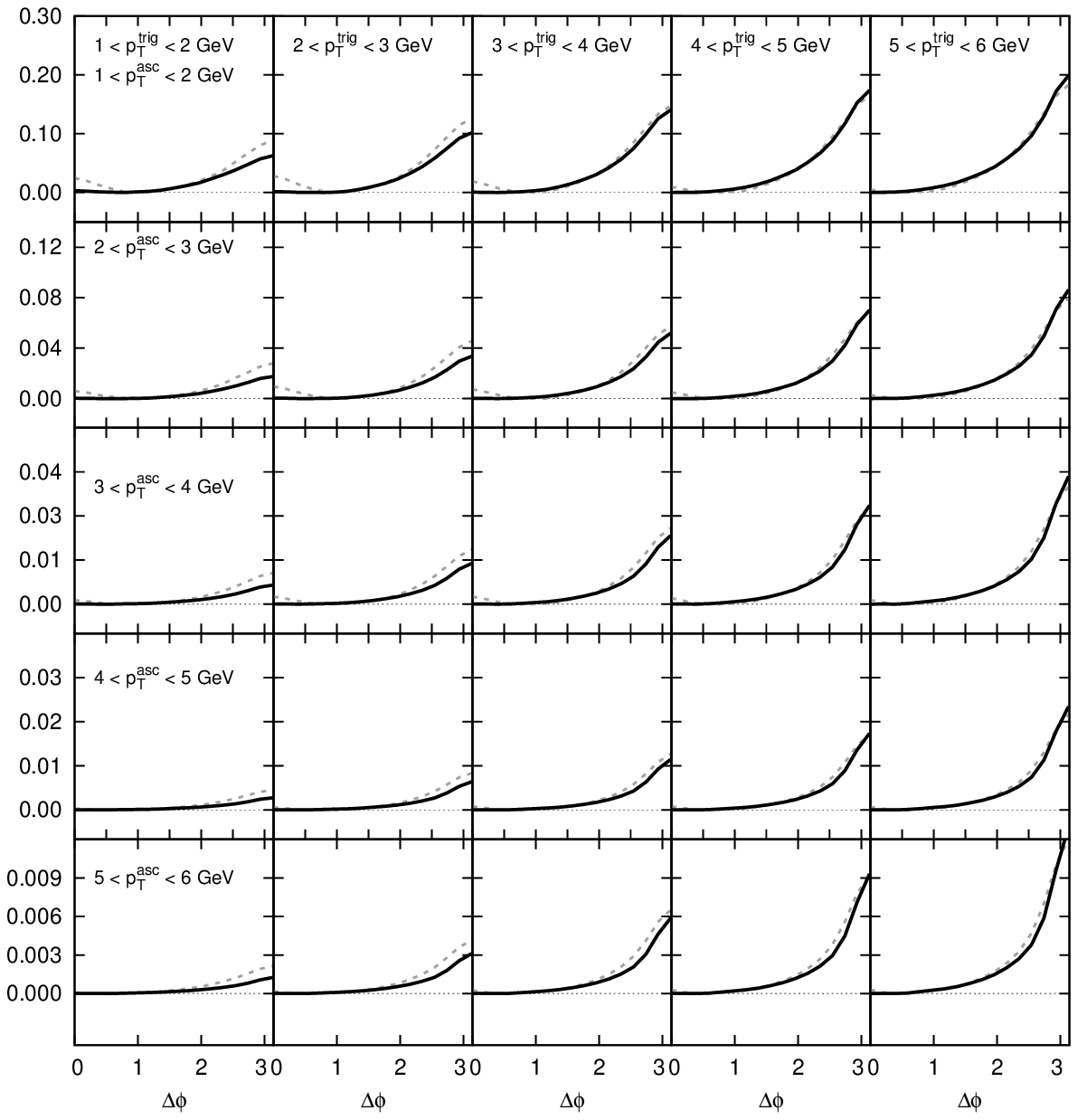}
\caption{BFKL prediction for the away-side yield in minimum bias p+Pb collisions at $\sqrt{s}=5$ TeV.
The gray curves show the corresponding results from central p+p.  The underlying
event has been subtracted from each bin.}
\label{fig:pPb_matrix}
\end{figure}
%%%%%%%%%%%%%%%%%%%%%%%%%%%%%%%%%%%%%%%%%%%%%%%%%%%%%%%%%%%%%%%%%%%%%%%%%

In Fig.~(\ref{fig:pp_matrix}), we compare the ``back-to-back" BFKL contribution
in Eq.~(\ref{eq:BFKL}), which dominates the awayside $\Delta\phi_{pq}$
distribution to data in the $\pasc$, $\ptrig$ matrix\footnote{This comparison also includes the awayside contribution from the Glasma graphs but, as discussed previously, this contribution is small. Because BFKL at the accuracy computed gives a {``small"} nearside anticollimation, this changes the Glasma $K$-factor from unity to $K=2.3$. See footnote~\ref{anti-coll}.}. The BFKL result gives a good agreement with awayside data in the entire matrix. This agreement can be contrasted with the result from the MRK expression in Eq.~(\ref{eq:MRK}), as well as the result of a computation in 
quasi-multi-Regge-kinematics (QMRK) that includes the $gg\rightarrow gggg$ matrix element in this kinematics~\cite{Fadin:1996zv,Leonidov:1999nc}. In Fig.~(\ref{fig:pPb_matrix}), we plot 
our prediction for the awayside di-hadron yield per unit $N_{\rm trig}$ in p+Pb collisions at $\sqrt{s}=5$ TeV compared to the same in central p+p. 

It will be very interesting to learn if our predictions are consistent with observations in the forthcoming p+Pb data from the LHC, and whether they can be distinguished from 
predictions in other related frameworks for di-hadron and jet spectra. With
regard to the latter, it will be interesting to extend our predictions to di-jet
production~\cite{Deak:2010gk,Kutak:2012rf} which has a larger rapidity acceptance at the LHC. Such tests of di-hadron spectra over the entire $\Delta\phi_{pq}$ range have the potential to provide fundamental insights into the nature of QCD bremsstrahlung and gluon saturation at high energies.

\section*{Acknowledgements}
 K.D.  and  R.V are  supported by the US Department of Energy under DOE Contract Nos.
DE-FG02-03ER41260 and DE-AC02-98CH10886 respectively. We would like to acknowledge useful conversations with 
Jochen Bartels, Dmitri Colferai, Francesco Hautmann and Samuel Wallon. 

\bibliography{biblio}{}

\end{document}